\documentclass[sigconf]{acmart}
\usepackage{hyperref}
\usepackage{makecell}
\usepackage{multirow}
\usepackage{enumitem}
\usepackage{url}
\usepackage{amsmath}
\usepackage{mdframed}
\usepackage{xcolor}
\usepackage{subcaption}
\usepackage{bm}
\usepackage{siunitx}
\DeclareSIUnit{\amperehour}{Ah}
 \usepackage{footnote}
\usepackage{amsmath}
\usepackage{mdframed}
\usepackage{xcolor}
\usepackage{subcaption}
\usepackage{bm}
\usepackage{siunitx}

\graphicspath{{figures/}}
\newmdenv[
    topline=true,
    bottomline=true,
    leftline=true,
    rightline=true,
    linecolor=blue,
    roundcorner=20pt,
    skipbelow=8pt,
    skipabove=10pt,
]{reviewbox}

\AtBeginDocument{%
  \providecommand\BibTeX{{%
    \normalfont B\kern-0.5em{\scshape i\kern-0.25em b}\kern-0.8em\TeX}}}

\newcommand\openGitHub{https://github.com/thuhci/OpenRing}

\copyrightyear{2025} 
\acmYear{2025} 
\acmConference[UbiComp Companion '25]{Companion of the 2025 ACM
International Joint Conference on Pervasive and Ubiquitous Computing}{October 12--16, 2025}{Espoo, Finland}
\acmBooktitle{Companion of the 2025 ACM International Joint Conference on
Pervasive and Ubiquitous Computing (UbiComp Companion '25), October
12--16, 2025, Espoo, Finland}\acmDOI{10.1145/3714394.3756252}
\acmISBN{979-8-4007-1477-1/2025/10}

\begin{document}

\title{$\tau$-Ring: A Smart Ring Platform for Multimodal Physiological and Behavioral Sensing}




\begin{abstract}
Smart rings have emerged as uniquely convenient devices for continuous physiological and behavioral sensing, offering unobtrusive, constant access to metrics such as heart rate, motion, and skin temperature. Yet most commercial solutions remain proprietary, hindering reproducibility and slowing innovation in wearable research. We introduce \textbf{$\tau$-Ring}, a commercial-ready platform that bridges this gap through: (i) \emph{accessible hardware} combining time-synchronized multi-channel PPG, 6-axis IMU, temperature sensing, NFC, and on-board storage; (ii) \emph{adjustable firmware} that lets researchers rapidly reconfigure sampling rates, power modes, and wireless protocols; and (iii) a fully \emph{open-source Android software suite} that supports both real-time streaming and $\ge\! $~8-hour offline logging. Together, these features enable out-of-the-box, reproducible acquisition of rich physiological and behavioral datasets, accelerating prototyping and standardizing experimentation. We validate the platform with demonstration studies in heart-rate monitoring and ring-based handwriting recognition. Source code is available at GitHub: \url{\openGitHub}.
\end{abstract}

\keywords{smart ring, open software, photoplethysmography, mobile health}

\begin{teaserfigure}
    \centering
    \includegraphics[width=1\linewidth]{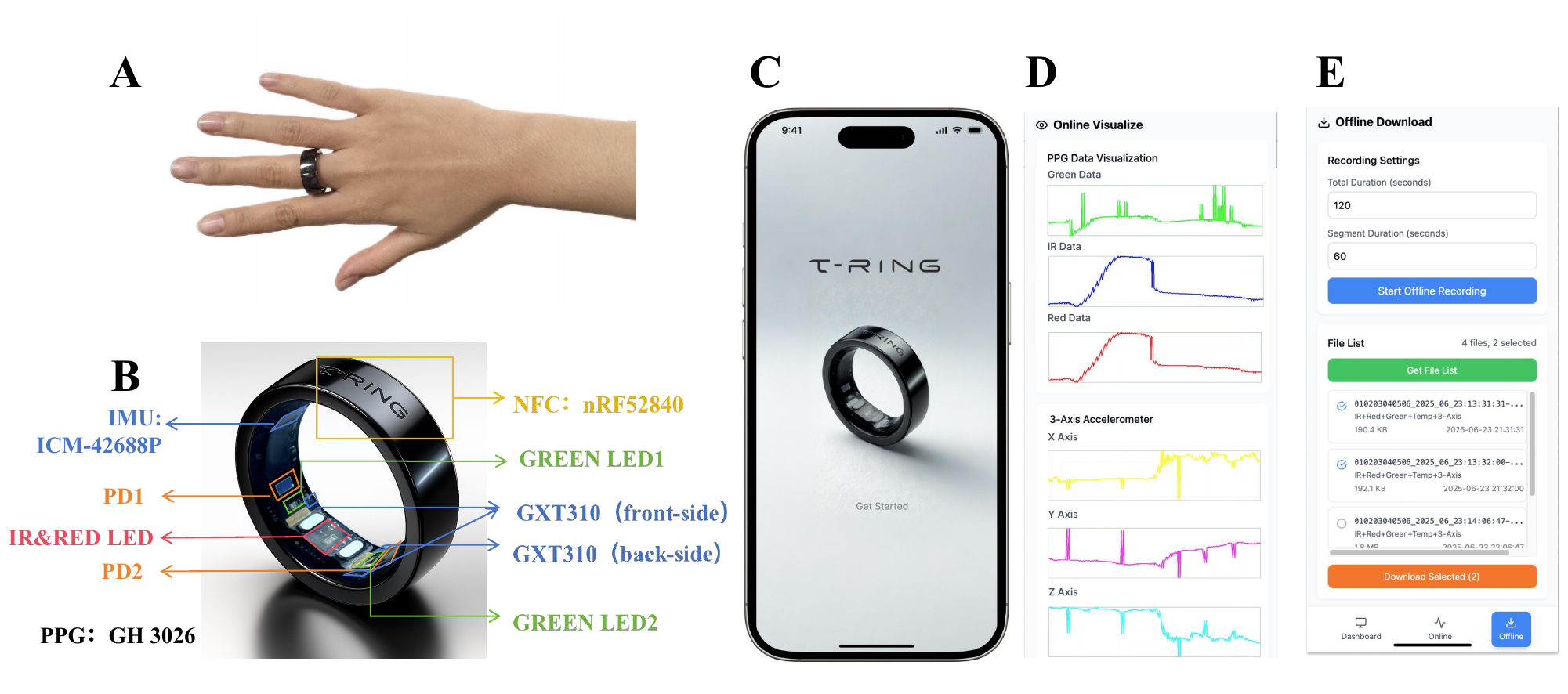}
\caption{\textbf{Overview of $\tau$-Ring.} 
(A) Worn on the finger for daily use. 
(B) Integrates PPG, IMU, TEMP, and NFC sensors. 
(C) Collection app main interface. 
(D) Real-time streaming with live visualization. 
(E) Offline logging for long-term data acquisition.}
    \label{fig:overview}
\end{teaserfigure}

\ccsdesc[500]{Human-centered computing~Ubiquitous and mobile computing}



\author{Jiankai Tang}
\authornote{Co-first authors}
\orcid{0009-0009-5388-4552}
\email{tjk24@mails.tsinghua.edu.cn}
\author{Zhe He}
\orcid{0000-0001-5874-1096}
\authornotemark[1]
\author{Mingyu Zhang}
\orcid{0009-0001-4918-0810}
\author{Wei Geng}
\orcid{0009-0000-5841-3498}
\affiliation{%
  \institution{Tsinghua University}
  \country{China}
}
\author{Chengchi Zhou}
\orcid{0009-0006-6620-3734}
\author{Weinan Shi}
\orcid{0000-0002-1351-9034}
\author{Yuanchun Shi}
\orcid{0000-0003-2273-6927}
\author{Yuntao Wang}
\authornote{Corresponding author}
\email{yuntaowang@tsinghua.edu.cn}
\orcid{0000-0002-4249-8893}
\affiliation{%
  \institution{Tsinghua University}
  \country{China}
}



\maketitle

\section{Introduction}
Smart rings have quickly grown in popularity due to their compact form-factor, attractive design, and convenience, driving significant market expansion from \$340.9 million in 2024 toward a projected \$2.53 billion by 2032 (CAGR 29.3\%)\footnote{\url{https://www.fortunebusinessinsights.com/smart-ring-market-111418}}. Commercial products such as the \textit{Oura Ring}\footnote{\url{https://ouraring.com/}} and the \textit{Samsung Galaxy Ring}\footnote{\url{https://www.samsung.com/us/rings/galaxy-ring/}} leverage photoplethysmography (PPG) and inertial sensors for physiological monitoring. However, these commercial devices expose only processed metrics with limited access to raw data and locked-down firmware, impeding algorithmic innovation and rigorous validation~\cite{koskimaki2018we,malakhatka2021monitoring,poongodi2022diagnosis,rajput2023assessment}.

While numerous research efforts have explored smart ring applications, from subtle interaction~\cite{rissanen2013subtle,tang2025exploring} to health monitoring~\cite{usman2019analyzing,halkola2019towards,tang2025dataset}, they predominantly rely on proprietary hardware platforms. This reliance creates significant barriers for reproducibility, cross-validation, and algorithmic standardization. With most researchers constructing custom one-off prototypes or repurposing commercial devices~\cite{tang2023alpha,liu2024summit}, the field struggles to establish benchmarks or facilitate direct comparisons between sensing methods~\cite{tang2024camera}, hindering general applicability of developed algorithms~\cite{wang2025computing,zhang2024earsavas}.

Recent efforts toward open smart ring platforms, such as OmniRing~\cite{zhou2023one}, have started addressing hardware accessibility. However, these initiatives often present significant reproduction barriers including complex fabrication requirements, limited documentation, and lack of open-source software support. Without comprehensive firmware and software ecosystems, even open hardware designs remain inaccessible to many researchers who lack specialized embedded systems expertise.

To address these challenges, we introduce \textbf{$\tau$-Ring}, a commercial-ready smart ring platform combining accessible hardware, adjustable firmware, and fully open-source software supporting both real-time streaming and long-term logging. By providing an out-of-the-box acquisition solution spanning all system layers, our platform enables researchers to rapidly prototype applications without extensive engineering investment. Through demonstration experiments in physiological sensing and motion recognition, we validate the platform's feasibility as a foundation for reproducible smart ring research.

Contributions of our work include:
\begin{enumerate}
    \item \textbf{$\tau$‑Ring platform}: a commercial‑ready, easy-to-use smart‑ring device with multi-channel PPG/IMU/TEMP sensing modalities and internal storage, enabling long-term continuous collection.
    \item \textbf{Adjustable firmware}: firmware allowing rapid re‑configuration of sensor, sampling, and communication parameters. 
    \item \textbf{Open software suite}: fully open (MIT-licence) mobile (Android) application offering online streaming and \(8\,\mathrm h\)+ continuous logging out‑of‑the‑box.
    \item \textbf{Demonstration experiments}: example case studies for HR monitoring during daily activities through PPG and ring handwriting recognition using the on‑board IMU.
\end{enumerate}

\section{Related Work}
Open-source wearables span EEG (\textsc{OpenBCI}), PPG (\textsc{Empatica E4}), audio (\textsc{OpenEarable}), and motion (\textsc{EmotiBit}); however, these platforms typically feature larger form factors and complex structures designed primarily for wrist or head mounting, limiting their wearability and comfort during extended use. In contrast, finger-ring platforms remain largely proprietary despite their compact size and natural integration with daily activities. Research prototypes such as DualRing~\cite{liang2021dualring}, EFRing~\cite{chen2023efring}, ThumbTrak~\cite{sun2021thumbtrak} and OmniRing~\cite{zhou2023one}, explore interaction or sensing modalities, but none provide a complete, commercial-ready hardware plus open firmware and software stack.

\begin{table*}[htp]
  \centering

  \begin{tabular}{lcccccc}
  \hline
  \textbf{Platform} & \textbf{Position} & \textbf{PPG} & \textbf{IMU} & \textbf{TEMP} & \textbf{Data Storage} & \textbf{Battery} \\
  \toprule
  \textbf{OpenBCI~\cite{kim2015openbci}} & Head-mounted & $\times$ & LIS3DH & $\times$ & \checkmark & \checkmark \\
  \textbf{EmotiBit~\cite{montgomery2023emotibit}} & Wrist/Head & MAX30101 & BMI160 & MLX90632 & $\times$ & \checkmark \\
  \textbf{OpenEarable 2.0~\cite{roeddiger2025openearable}} & Earbud & MAXM86161& BMX160 & MLX90632 & \checkmark & \checkmark \\
  \textbf{OmniRing~\cite{zhou2023one}} & Finger & MAX30101& ICM-20948 & $\times$ & $\times$ & \checkmark \\
  \midrule
  \textbf{$\tau$-Ring(Ours)} & \textbf{Finger} & \textbf{GH3026} & \textbf{ICM-42688P} & \textbf{GXT310} & \checkmark & \checkmark \\
  \bottomrule
  \end{tabular}
  
  \caption{Comparison of open-source research wearable devices and their sensor components. PPG = photoplethysmography, IMU = inertial measurement unit, TEMP = temperature sensor.}
    \label{tab:related}
\end{table*}

\subsection{Ring-based Technologies}
In recent years, ring-based technologies have experienced rapid development. A recent comprehensive survey~\cite{wang2025computing} reviewed 206 smart ring-related studies, highlighting a growing trend in the number of publications in this area. We conducted a detailed analysis of these studies and quantified the use of IMU, PPG, and temperature sensors. Specifically, 62 studies employed IMU sensors, 7 used PPG, 7 combined PPG with temperature sensing, 3 combined PPG with IMU, and 7 integrated PPG, IMU, and temperature sensors. In total, 86 studies utilized at least one of these three sensors or their combinations. In principle, all these studies could be replicated and validated using the $\tau$-Ring platform.

\textbf{Interactions.} Due to their portability and always-available nature, smart rings have been widely explored in interaction research. For example, Tickle~\cite{wolf2013tickle} utilized an IMU to detect tapping gestures on handheld objects. Dualring~\cite{liang2021dualring} employed two IMUs to recognize thumb-to-index-finger taps and swipes, enabling cursor control. ThumbRing~\cite{boldu2018thumb} used IMUs worn on the thumb and back of the hand to detect thumb taps on different finger joints, achieving an accuracy of up to 92.3\%. WashRing~\cite{xu2022washring} leveraged a ring-mounted IMU to classify various handwashing gestures with 97.8\% accuracy. Touch+Finger~\cite{lim2018touch+} enhanced multi-finger gestures with an IMU on a ring worn on the middle finger, achieving over 99\% accuracy. Ready, Steady, Touch~\cite{shi2020ready} used a single IMU ring to detect finger contacts with different surfaces, reaching over 95\% accuracy and even distinguishing between surface materials. MouseRing~\cite{shen2024mousering} enabled virtual touchpad interaction on arbitrary surfaces using one or two IMU rings, achieving performance comparable to a real touchpad in a Fitts’ Law test. Smart rings have also been used for text input. WritingRing~\cite{writingring} reconstructed planar handwriting trajectories using a single IMU ring, reaching 88.7\% character accuracy and 68.2\% word accuracy. QwertyRing~\cite{gu2020qwertyring} enabled word-level text entry using an IMU ring worn on the index finger, achieving a peak text entry speed of 20.59 WPM.

\textbf{Physiological sensing.} Smart rings equipped with PPG and temperature sensors are commonly used for physiological monitoring. For instance, ~\citet{boukhayma2021ring} proposed an optimized PPG design for enhanced heart rate sensing, achieving 97.87\% accuracy. SensoRing~\cite{mahmud2018sensoring} and OmniRing~\cite{zhou2023one} integrated IMU, temperature, and PPG sensors, reporting mean absolute errors of 3 bpm and 8 bpm, respectively, for heart rate monitoring. ~\citet{santos2022use} evaluated a ring-based pulse oximeter and emphasized the significant impact of motion artifacts on SpO2 measurements. ~\citet{paliakaite2021blood} compared PPG-based blood pressure estimation between the finger and the wrist. The mean ± standard deviation of the differences between reference and estimated systolic pressures, as well as the mean absolute error (MAE), were: 0.47 ± 10.44 mmHg and 7.78 mmHg for finger PPG; and 1.05 ± 12.86 mmHg and 9.69 mmHg for wrist PPG, suggesting that finger-based models more effectively captured BP trends during cold pressor tests. ~\citet{wongtaweesup2023using} evaluated the Wellue O2 ring using overnight SpO2 data to classify the severity of obstructive sleep apnea. Compared to the gold standard polysomnography (PSG), the detection accuracy was around 65\%, and notably outperformed smartwatch-based methods.

While many other smart ring projects are not listed here, in theory, almost all the aforementioned studies could be replicated using the $\tau$-Ring platform. Traditionally, such research required custom hardware development or the use of commercial products, both of which introduced substantial overhead. $\tau$-Ring addresses these challenges by providing readily accessible hardware and open-source software with zero setup barriers, significantly lowering the cost of data collection. Furthermore, previous studies often relied on disparate platforms and did not release their datasets, making reproducibility and reuse difficult—issues that $\tau$-Ring aims to systematically address.

\subsection{Open-source wearables}\label{sec:related_open}

Open-source and research-grade wearable platforms have been instrumental in lowering the barrier to physiological sensing studies.  Table~\ref{tab:related} summarises five representative devices that are widely cited in the community, together with our proposed finger-worn \emph{$\tau$-Ring}.  
The commercial-research \emph{Empatica E4} (wrist, 2014) opened large-scale stress-monitoring studies by streaming raw PPG and motion to on-board flash, but its closed SDK hides LED-drive and ADC settings, limiting reproducibility. \emph{OpenBCI} (head-mount, 2015)~\cite{kim2015openbci} democratised EEG with open schematics, yet lacked integrated PPG or temperature sensing, so users often stacked extra boards—penalising weight and the \SI{500}{\milli\amperehour} battery budget.

 \emph{EmotiBit}~\cite{montgomery2023emotibit} popularised a modular “strap-anywhere” philosophy: a Feather-compatible mainboard streams raw multi-wavelength PPG, nine-axis motion, skin temperature and EDA over BLE while buffering to a built-in microSD.  Its open firmware and code make it attractive for rapid prototyping, yet the sandwich form factor (board\,+\,battery\,+\,sensor strip) remains bulky for long-term ambulatory studies.  
 \emph{OpenEarable 2.0}~\cite{roeddiger2025openearable} moves sensing to the auricle, exploiting the ear canal’s mechanical stability for high-fidelity audio, PPG and IMU data. The platform introduces an ingenious dual-board design (rigid main PCB + flexbile sensor PCB) but still relies on a CR2032 (\SI{108}{\milli\amperehour}) coin cell and plug-in microSD, limiting multi-day capture unless the user recharges.  

Finally, the finger-mounted  \emph{OmniRing}~\cite{zhou2023one} pioneered high-density IMU-PPG fusion for gesture and heart rate tracking, but its proprietary firmware restricts access to intermediate ADC values and adaptive LED current control. The absence of temperature sensors limits its use in thermoregulation studies, and the
complex manufacture (e.g. PCB and sealing-in) may hinder adoption by the wider community.
Moreover, OmniRing lacks on-board flash, streaming every packet via BLE---a bottleneck in crowded \SI{2.4}{\giga\hertz} spectra.

\noindent\textbf{Motivation for $\tau$-Ring.}  Three key observations inform our design: (\emph{i}) compact, comfortable form factors are essential for enabling continuous daily wear in real-world environments, (\emph{ii}) accessible, ready-to-use hardware platforms accelerate research adoption without manufacturing efforts, and (\emph{iii}) achieving multi-modal, highly synchronized, long-duration data collection remains critical for comprehensive physiological monitoring. Our $\tau$-Ring addresses all three by integrating a sub-\SI{1}{\milli\metre} flex PCB around the finger, coupling a high-sensitivity three-LED PPG (GH3026), an industrial-grade low-noise IMU (ICM-42688P) and a miniature digital thermistor (GXT310) with a \SI{128}{\mega\byte} storage flash for untethered, lossless logging.  Powered by a \SI{15}{\milli\amperehour} curved Li-Po battery, the device supports \textbf{8 hours} of continuous three-modality recording at \SI{100}{\hertz}, and can last longer with optimized sampling rates and reduced LED intensity.

\begin{figure*}
  \includegraphics[width=0.9\linewidth]{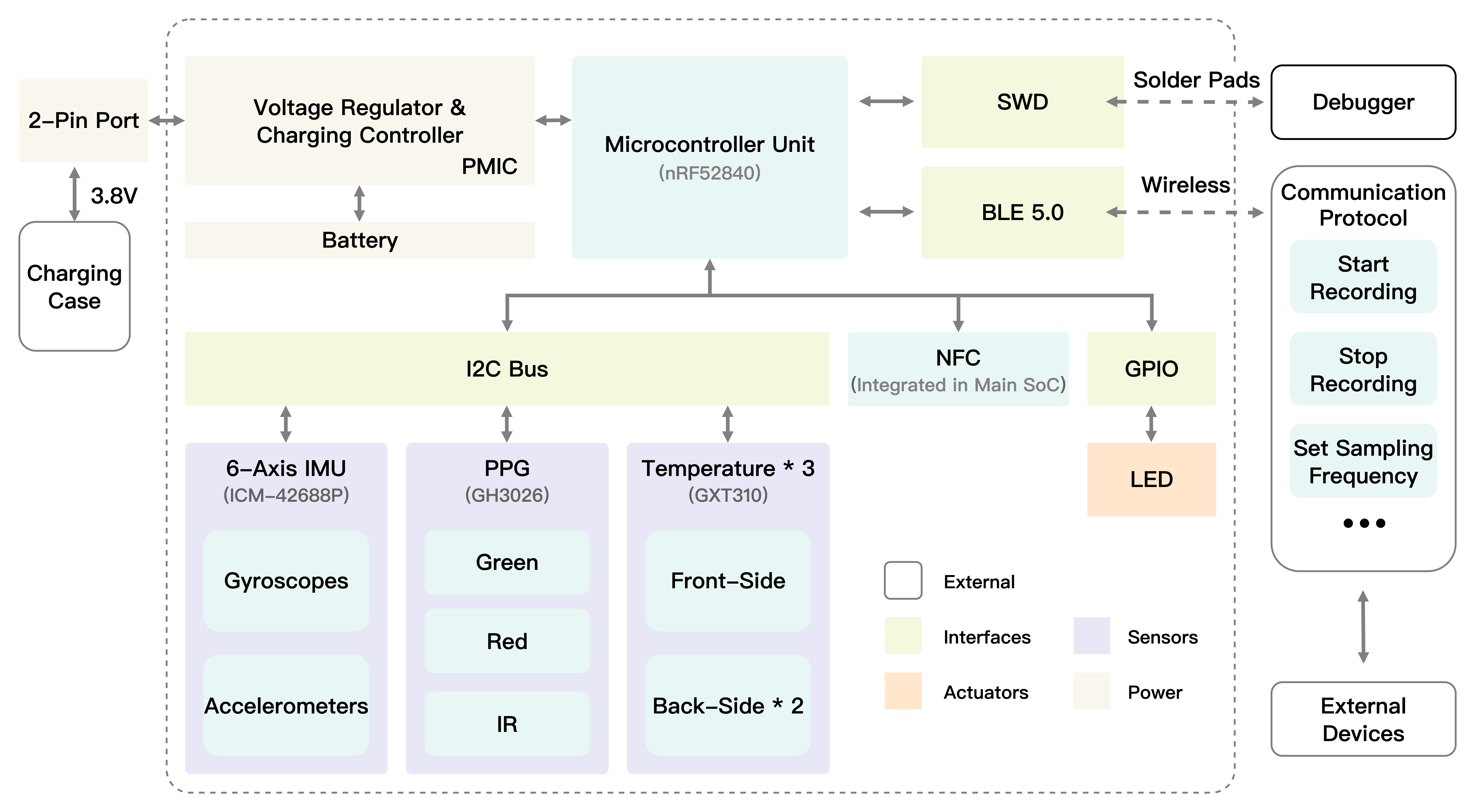}
  \caption{An overview of $\tau$-Ring's system architecture, encompassing crucial aspects such as hardware selection, data links, and communication protocols.}
  \label{fig:firmware}
\end{figure*}

\begin{figure*}[t]
  \centering
  \includegraphics[width=0.9\linewidth]{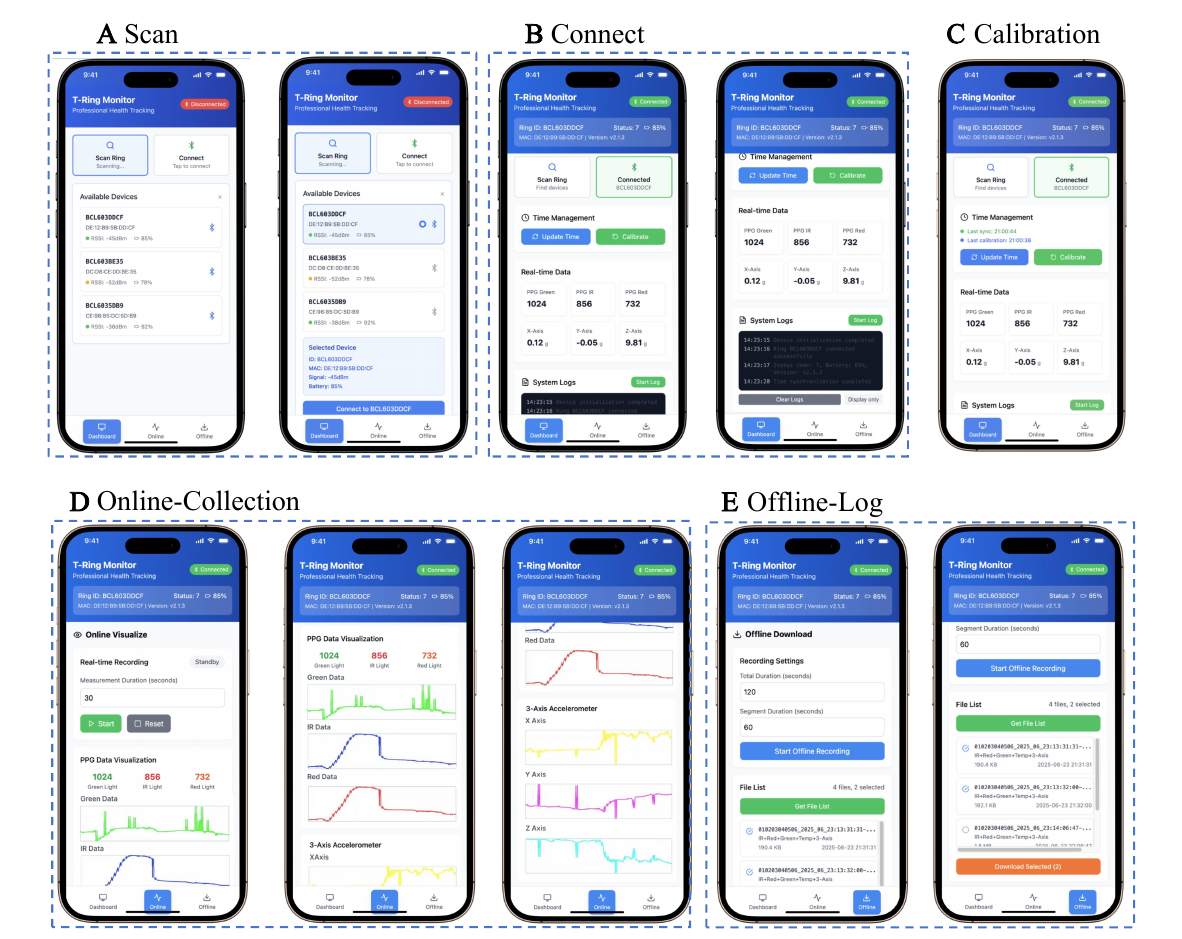}
  \caption{\textbf{Software UI and functionality.}
  (A) Scan nearby rings and list metadata.
  (B) Connect and display device details.
  (C) One-click clock calibration with drift feedback.
  (D) Online collection with live charts and annotations.
  (E) Configure, list and retrieve offline logs for long-term studies.}
  \label{fig:software}
\end{figure*}

\section{Method}

\subsection{Hardware Overview}\label{sec:hw_overview}

The $\tau$-Ring (Fig.~\ref{fig:overview}) is architected around an \textsc{nRF52840}\footnote{\url{https://www.nordicsemi.com/Products/nRF52840}} SoC that combines a 64\,MHz Cortex-M4F, 1\,MB on-chip flash and 256\,kB RAM with BLE\,5.0 and 2.4\,GHz proprietary radio. 

\textbf{(A) Optical stack.}  The three-wavelength PPG engine is a \textsc{GH3026}\footnote{\url{https://www.goodix.com/en/product/sensors/health_sensors/gh3026}} with individual LED current drivers (0-200\,mA, 8-bit), a 24-bit TIA ADC and ambient light cancellation.  We expose LED amplitude, pulse width and sampling frequency via adjustable firmware so researchers can replicate illumination protocols across studies. 

\textbf{(B) Inertial subsystem.}  We pair the optical stack with an \textsc{ICM-42688P}\footnote{\url{https://invensense.tdk.com/products/motion-tracking/6-axis/icm-42688-p/}} IMU (±16\,g, ±4000\,dps, \SI{0.88}{\milli\ampere} ODR=\SI{1}{\kilo\hertz}) whose integrated programmable digital filters align precisely with PPG sampling edges through the SoC’s DPPI hardware triggers.  Time-domain jitter across modalities is thus capped at \(\le\)\SI{8}{\micro\second}.

\textbf{(C) Thermal sensing.}  Finger skin temperature is captured via a \textsc{GXT310}\footnote{\url{https://doc.chipmall.com/datasheet/rev_2412070020_gxcas-gxt310t0_c51592898.pdf}} digital thermistor (±0.1\,$^{\circ}$C, \SI{16}{\bit}), positioned in two points inside the ring and one point outside the ring.

\textbf{(D) Storage and Battery.}  A \SI{128}{\mega\byte} storage flash (GD55B01GFYIGR) buffers raw data at up to \SI{100}{\hertz} across all three modalities, enabling multi-day logging without radio interference.  The ring is powered by a \SI{15}{\milli\amperehour} Li-Po cell (Constant Li-156825), which provides up to \SI{8}{\hour} of continuous operation at \SI{100}{\hertz} across all sensors.

In sum, $\tau$-Ring delivers research-grade synchronised PPG-IMU-TEMP streams in a \SI{2.4}{\gram} finger-form factor while achieving multi-day autonomy and fully offline logging—bridging the gap between open wearables and the usability demanded by longitudinal health studies.

\subsection{Firmware Architecture}
The $\tau$-Ring runs our custom-developed firmware (written in C). The system architecture is shown in Fig. \ref{fig:firmware}. It provides the following primary functionalities:

\textbf{(A) Data Acquisition}: The system collects data from various sensors in a polling-based manner, allowing for the simultaneous sampling of all sensors at a predetermined frequency.

\textbf{(B) Data Storage and Transmission}: It offers the flexibility to either store the collected data on the built-in flash memory or transmit it in real-time to an external device via BLE. During real-time transmission, data is sent out in packets every 50 ms, which is sufficient to support a live demonstration.

\textbf{(C) Remote Control and Configuration}: The device firmware is designed to be \textbf{adjustable}. It communicates with external devices via BLE, enabling them to send commands for control and configuration. Users can select between real-time transmission and offline storage, enable or disable specific sensors, set the sampling frequency for individual sensors, and perform time calibration and schedule future data acquisition tasks.

\subsection{Software Suite}
Our Android-based companion app unifies the entire acquisition workflow in a single, tabbed interface (Fig.~\ref{fig:software}).  Each tab corresponds to one of the suite’s five core functions, allowing both novices and power users to move seamlessly from initial discovery to long-term field deployments.

\textbf{(A) Scan nearby rings and list metadata} (Fig.~\ref{fig:software}A).  
At launch the app issues a Bluetooth \emph{extended scan} and instantly populates a sortable list of all discovered rings.  Alongside the device name we display RSSI, battery level, and advertised firmware version, so researchers can pick the strongest, most up-to-date unit even in radio-dense gyms or clinics.

\textbf{(B) Connect and display device details} (Fig.~\ref{fig:software}B).  
Tapping a row establishes an encrypted GATT link and pivots to a dashboard that exposes MAC address, currently mounted sensors, sampling rates and remaining flash capacity.  A colour-coded health indicator warns of low battery or sensor faults, reducing setup errors before long experiments.

\textbf{(C) One-click clock calibration with drift feedback} (Fig.~\ref{fig:software}C).  
Pressing “Calibrate” sends a 32-bit UNIX epoch to the ring, measures round-trip latency and computes instantaneous drift.  If the offset exceeds \SI{1}{\second}, the on-board RTC is trimmed and the test repeats until convergence.  The final offset and number of iterations are logged for audit-ready provenance when synchronising multi-modal datasets.

\textbf{(D) Online collection with live charts and annotations} (Fig.~\ref{fig:software}D).  
Entering \emph{Online} mode turns the phone into a miniature oscilloscope: LED-level PPG and triaxial accelerometer streams render at \SI{30}{\hertz} on a WebGL canvas, while numeric widgets update heart-rate and activity counts in real time.  Users can annotate the session with custom tags (e.g. “walking”, “resting”) that are timestamped and stored alongside the data, enabling post-hoc analysis of activity-specific physiological responses. 

\textbf{(E) Configure, list and retrieve offline logs for long-term studies} (Fig.~\ref{fig:software}E).  
For studies that outlast BLE range, switching to \emph{Offline} arms the ring’s internal scheduler: users set total duration and segment length, after which the radio sleeps to conserve power.  A progress bar tracks flash occupancy, and \texttt{Get File List} enumerates recordings with start time, size and CRC.  Selected segments are downloaded via BLE \texttt{\small L2CAP CoC} at \SI{128}{\kilo\bit\per\second}, ensuring rapid data offload without interrupting the next participant’s session.

\subsection{Demonstration Experiments}
\paragraph{Heart‑rate monitoring.} A study (n = 34)~\cite{tang2025dataset} benchmarked the $\tau$-Ring's IR PPG performance against commercial-grade devices (Oura Ring Gen 3 and Samsung Ring) during daily activities. Participants engaged in various activities, including sitting, standing, and walking, while wearing FDA-approved oximeter CMS50D+ as a reference. The results showed that the $\tau$-Ring achieved better performance than the Oura Ring and Samsung Ring compared with medical reference, with a lowest mean absolute error (MAE) of 5.18 BPM in stationary scenarios. 

\paragraph{Ring handwriting.} Prior work has investigated the use of Inertial Measurement Units (IMUs) integrated into ring-form-factor devices for hand gesture recognition. Given the physical consistency of IMU data, these established methodologies can be replicated easily on the $\tau$-Ring, which can also serve as a platform for future explorations in this domain.

To demonstrate this capability, we developed an application inspired by WritingRing ~\cite{writingring}, which reconstructs the user's writing trajectory to recognize English letters and words. In our implementation, a user wears the $\tau$-Ring on their right index finger and performs writing gestures on a flat surface, such as a tabletop. A Temporal Convolutional Network with Long Short-Term Memory (TCN-LSTM) model was employed to reconstruct the writing trajectory from the IMU data. The reconstructed trajectory was subsequently input into the Google Input Method Editors (Google IME)\footnote{https://www.google.com/inputtools/} for letter and word recognition.

We conducted a user study with three participants to evaluate the system. Following a brief warm-up session of approximately five minutes, each participant was instructed to write the complete English alphabet (both uppercase and lowercase) and a set of 20 randomly selected English words from a list of the 3000 most common English words ~\cite{google10000}. The evaluation yielded a recognition accuracy of 88.5\% for individual letters, and an accuracy of 55.0\% for words. These results are comparable to those reported in the original WritingRing paper.

This evaluation demonstrates that the hardware and firmware design of the $\tau$-Ring is robust and capable of reproducing the results of previous IMU-based ring research. Furthermore, it underscores the potential of the $\tau$-Ring to enhance the reproducibility of studies within the field of ring interaction technology.

\section{Discussion}
The $\tau$-Ring platform addresses critical gaps in wearable research infrastructure by providing a complete, accessible solution spanning hardware, firmware, and software. Unlike commercial devices that restrict access to raw data or research prototypes requiring specialized fabrication expertise, our platform enables immediate deployment with minimal technical barriers. This accessibility facilitates reproducible research—a persistent challenge in wearable computing where proprietary hardware often prevents algorithm validation across studies.

The platform's modular architecture offers significant advantages for both novice and experienced researchers. Beginners can leverage pre-built profiles for common applications like heart rate monitoring, while advanced users can directly modify sensor parameters and processing pipelines without rebuilding firmware. This flexibility, combined with comprehensive documentation and open-source repositories, creates opportunities for community-driven innovation and standardization of ring-based sensing algorithms.

Furthermore, $\tau$-Ring's ability to support both online streaming and offline logging addresses practical challenges in longitudinal studies. The 8+ hour continuous logging capability enables data collection throughout daily activities without requiring constant smartphone proximity or researcher supervision. By providing synchronized multi-channel data acquisition in a comfortable, finger-worn form factor, the platform enables exploration of previously challenging research questions around continuous physiological monitoring in naturalistic settings.

\section{Limitation and Future Work}

Despite its capabilities, the current $\tau$-Ring implementation faces several limitations. The 15\,mAh battery, while sufficient for continuous day-long studies, requires recharging for multi-day continuous monitoring. Future iterations could explore higher-capacity curved batteries or more aggressive power management strategies. In addition, although we provide open-source data acquisition software and firmware interfaces, full hardware openness is limited by manufacturing constraints and supply chain dependencies—particularly in areas such as flexible PCB fabrication and sensor integration.

From a sensing perspective, our platform currently lacks electrodermal activity (EDA) sensing, which would enable stress and emotional arousal monitoring. Integrating miniaturized EDA electrodes presents mechanical challenges that future work could address. Additionally, the companion application is currently Android-only, limiting deployment in iOS-based research environments.

Future work will focus on three primary directions: (i) enhancing the hardware platform with additional sensing modalities and improved battery life; (ii) expanding software compatibility to iOS and enabling more sophisticated on-device processing for power-constrained scenarios; and (iii) developing a community repository of validated signal processing algorithms specifically optimized for ring form factors. We also plan to explore interoperability with other open wearable platforms to support multi-device sensing protocols for comprehensive physiological monitoring.

\section{Conclusion}
We presented $\tau$-Ring, a smart ring platform that addresses significant barriers in wearable research. By integrating research-grade hardware with adjustable firmware and open-source software, $\tau$-Ring enables reproducible experiments without specialized engineering expertise. The platform's synchronized multichannel sensing capabilities, coupled with both real-time streaming and 8+ hour offline logging, support a wide range of applications—from continuous physiological monitoring to motion-based interaction. Our demonstrations of heart rate monitoring and handwriting recognition validate the platform's versatility. By providing this complete data acquisition solution to the OpenWearable community, we aim to accelerate innovation in ring-based sensing and establish more standardized, comparable methodologies for future wearable computing research.

\begin{acks}
This work is supported by the National Key R\&D Program of China under Grant No. 2024YFB4505500 \& 2024YFB4505503, the National Natural Science Foundation of China under Grant No. 62366043 \& 62472244, the foundation of National Key Laboratory of Human Factors Engineering under Grant No. HFNKL2024W06, the Tsinghua University Initiative Scientific Research Program under Grant No.
20257020004, and Qinghai University Research Ability Enhancement Project under Grant No. 2025KTSA05.
\end{acks}

\bibliographystyle{ACM-Reference-Format}
\bibliography{openwearable}


\end{document}